\documentclass[9pt,twocolumn,aps,pre]{revtex4}
\usepackage{subfigure}
\usepackage{graphicx}
\usepackage[utf8]{inputenc}

\usepackage{float}
\usepackage{amssymb}

\begin{document}

\sloppy
\newcommand{\rs}{\begin{equation}}
\newcommand{\re}{\end{equation}}
\newcommand{\eas}{\begin{eqnarray}}
\newcommand{\eae}{\end{eqnarray}}

\renewcommand{\subfigtopskip}{0.1cm}
\renewcommand{\subfigcapskip}{0cm}
\renewcommand{\subfigbottomskip}{0cm}
\renewcommand{\subfigcapmargin}{0cm}

\title{Long-range effects on superdiffusive solitons in anharmonic 
chains}
\author{C. Brunhuber}    
\email{Christian.Brunhuber@uni-bayreuth.de}
\author{F.G. Mertens}
\affiliation{Physikalisches Institut, Universit\"at Bayreuth, D-95440 Bayreuth, Germany}
\author{Y. Gaididei}
\affiliation{Bogolyubov Institute for Theoretical Physics, 03143 Kiev, Ukraine}

\begin{abstract}

Studies on thermal diffusion of lattice solitons in Fermi-Pasta-Ulam (FPU)-like lattices
were recently generalized to the
case of dispersive long-range interactions (LRI) of the Kac-Baker form.
The position variance of the soliton shows a stronger than linear
time-dependence (superdiffusion) as found earlier for lattice solitons on FPU chains with
nearest neighbour interactions (NNI). 
In contrast to the NNI case where the position variance at moderate soliton
velocities has a considerable
linear time-dependence (normal diffusion), the solitons
with LRI are dominated by a superdiffusive mechanism where the
position variance mainly depends quadratic and cubic on time.   
Since the superdiffusion seems to be generic for
nontopological solitons, we want to illuminate the role of the soliton shape on
the superdiffusive mechanism.
Therefore, we concentrate on a FPU-like lattice with a certain class of power-law long-range interactions where the solitons have algebraic tails instead of 
exponential tails in the case of FPU-type interactions (with or without Kac-Baker LRI).
A collective variable (CV) approach in the continuum approximation of the system
leads to stochastic integro-differential equations which can be reduced to Langevin-type equations for the 
CV position and width. We are able to derive an analytical result for the soliton diffusion which agrees well with 
the simulations of the discrete system. Despite of structurally similar
Langevin systems for the two soliton types, the
algebraic solitons reach the superdiffusive long-time limit with a
characteristic $t^{1.5}$ time-dependence much faster than
exponential solitons. The soliton shape determines the diffusion constant in the
long-time limit that is approximately a factor of $\pi$ smaller for algebraic
solitons.

\end{abstract}

\maketitle
\section{Introduction}
Nonlinear excitations like solitons and discrete breathers have been drawing great attention over recent years.
They present very robust solutions of nonlinear partial differential equations and nonlinear lattice models which are often used to describe a rather broad set of 
physical systems \cite{scott,encyclopedia,dauxois1}. Nonlinear lattices like Klein-Gordon (KG), Fermi-Pasta-Ulam or the discrete nonlinear Schroedinger (DNLS) 
lattice (and their counterparts in the continuum approximation) present all to
often a strong approximation of the system of interest and it is sometimes not clear 
to what extend solitons or discrete breathers are also relevant in more realistic models.  
In the case of biomolecules, the aspiration to highlight their basic functionalities with computer simulations is a challenging task for the computational science nowadays. If we aim to understand principally the role of
nonlinear excitations in
biomolecules, we have to investigate simpler models than the numerical ab initio
calculations. Often used in this context is the Davydov model \cite{davydov,scott92} for energy and charge transport in proteins or the
Peyrard-Bishop-Dauxois (PBD) model which describes the melting and
the denaturation of DNA \cite{peyrard04,yakushevic01}.   
There are convincing evidences that some features of nonlinear excitations
in this simplified models are relevant in explaining the functionality of biomolecules. Recently, pump-probe
measurements \cite{edler02} showed that the lifetime of NH stretching bands in the model protein ACN is (with about 18 ps) in good agreement with numerical calculations of the Davydov model
\cite{cruzeiro97}. The observation that the PBD model succeeds in determining the thermally induced openings of the DNA strand at functionally relevant sites
for the DNA transcription \cite{choi04} is also an example for the relevance of nonlinear lattice models in biology. \\
Nowadays, many physicists who work in this field try to 
extend and improve the basic models in order to make them more realistic. The spatial structure of biomolecules, thermal fluctuations, damping and  long-range effects stemming from Coulomb or dipole-dipole interactions 
have certainly a great influence on the nonlinear excitations and the statistics of the system. 
It is known that long-range effects can change the features of solitons and discrete breathers qualitatively when the LRI exceed some critical
value \cite{gaididei97,flach98}. In FPU-like 
chains and nonlinear Schr\"odinger models, they give rise to new types of solitons which can coexist at the same value of the spectral parameter \cite{christiansen01}. Controlled
switching between such soliton states was recognized as a possible mechanism for energy transport and storage in biomolecules \cite{mingaleev99}.  
In the past, nonlinear KG lattices with LRI were frequently studied in investigations of a number of physical phenomena
such as dislocations in solids, charge density waves, absorbed layers of atoms or domain walls in ferromagnets and ferroelectrics (see references in \cite{braun98,bonart97,roessler2000}).
The effect of thermal fluctuations and LRI are usually regarded separately
\cite{cruzeiro97,cruzeiro98,rasmussen98} because of their complexity. Nevertheless it is known that LRI
can have very interesting effects on the thermodynamics of many different physical systems \cite{dauxois,brunhuber06}.
\\
 The Kac-Baker form presents a spatially exponential coupling between different particles
which 
is quite often chosen for the inclusion
of LRI in nonlinear lattice models \cite{gaididei97,brunhuber06,mingaleev00}.
In the recent publication \cite{brunhuber06i} lattice solitons on FPU-like chain with Kac-Baker LRI in the presence of a thermal reservoir were investigated.
In the continuum approximation of the system and with a collective variable (CV) approach, it was possible to derive a Langevin set for the soliton position, similar to the 
solitons on FPU chains \cite{arevalo03}. The solitons on FPU chains (with or without Kac-Baker LRI) show superdiffusive behaviour. The position variance of the
soliton shows besides the linear term in the time-dependence also quadratic and cubic terms. In \cite{brunhuber06i}, it was demonstrated, that the diffusion mechanism for NNI and
for additional LRI are quite similar because the same CV ansatz was used. 
In the case of Kac-Baker LRI, this approximation is valid unless the soliton velocities do not reach the critical velocity, where the soliton 
begins to develop a cusp \cite{gaididei97}.
\\
In order to check the influence of the soliton shape on the diffusion of the soliton, we choose a power-law coupling. This leads in the continuum approximation (CA) 
to a Benjamin-Ono (BO) equation, which is known to posses algebraic soliton
solutions \cite{mingaleev98}. The first studies in that direction date back to
Ishimori \cite{ishimori82} who
studied anharmonic chains with Lennard-Jones ($2n$,$n$) intermolecular potential and showed that the dynamics is governed by the Benjamin-Ono equation in the case $n=2$ or
by the Korteweg-de Vries equation for $n \ge 4$.   
In the case of cubic and quartic nearest neighbour interactions, a reductive perturbation
method yields a KdV+mKdV and Benjamin-Ono equation wich was shown to possess exact
nonsingular rational solutions \cite{castea98}.   
\section{The Model}
Our model is a one-dimensional chain of equally spaced particles of mass m ($m=1$) with an
interatomic spacing $a$ ($a=1$). We denote the displacement of the particle $n$ from its
equilibrium position as $u_n$ (absolute displacement coordinates) and the difference $w_n=u_{n+1}-u_n$ as relative displacement coordinates. 
The potential consists of a part $U_{NN}$ similar to the original potential chosen by Fermi, Pasta and Ulam and a long-range potential $U_{LR}$ with a power-law 
dependence of the harmonic coupling constant $J_{nm}$ 
\eas
T=\frac{1}{2} \sum_{n=1}^N \bigg(\frac{d u_n(t)}{d t} \bigg)^2
\\
U_{NN}=\sum_{n=1}^{N} V(u_{n+1}(t)-u_n(t))
\\
V(r)=\frac{r^2}{2}-\frac{r^3}{3}
\\
U_{LR}=\frac{1}{2} \sum_n \sum_{m\neq n}  \frac{J}{\mid m-n \mid^s} (u_n-u_m)^2 ~~.
\eae
The model covers the physical situation of dipole-dipole ($s=5$) and
Coulomb interactions ($s=3$) between the particles on the chain, if we restrict ourselves to small relative displacements. 
The equation of motions in relative displacement coordinates can be obtained from the
Lagrangian of the system $L=T-U_{NN}-U_{LR}$
\eas
\ddot{w}_i(t)&=&V^{\prime}(w_{i+1})-2 V^{\prime}(w_i)+V^{\prime}(w_{i-1})\nonumber \\&-&J
\sum_{m \neq 0} \frac{(w_i-w_{i+m})}{\mid m \mid^s}~.
\label{BONOdisceq}
\eae  
When we want to consider the effect of a thermal bath, we can add damping and noise terms to
the original equation (\ref{BONOdisceq}) such that they fulfill the
fluctuation-dissipation theorem ($D^{hy}=2 \nu_{hy} T$ with $k_B$ set to unity) \cite{arevalo03}.

\eas
\ddot{w}_i(t)&=&V^{\prime}(w_{i+1})-2 V^{\prime}(w_i)+V^{\prime}(w_{i-1})\nonumber \\&-&J
\sum_{m \neq 0} \frac{(w_i-w_{i+m})}{\mid m \mid^s}
+ \nu_{hy}(\dot{w}_{i+1}-2 \dot{w}_i+\dot{w}_{i-1})\nonumber \\&+&\sqrt{D^{hy}}(\xi_{i+1}(t)-\xi_{i})=0~.
\label{BONOdisceq1}
\eae
We choose hydrodynamical damping which depends on the relative displacement velocities of the
particles and presents an intrinsic damping mechanism of the system. The widely used Stokes
damping is not appropriate for pulse solitons because one obtains an imaginary dispersion
relation the long-wave region of the Fourier
spectrum what causes deformations of the soliton \cite{arevalo02}.     
We go to the continuum limit [$w_i(t) \rightarrow w(x,t)$ , $\xi_i(t) \rightarrow \xi(x,t)$ , $f(x+m,t) \rightarrow e^{m
\partial_x} f(x,t)$] and find the partial differential equation (PDE) 
\eas
\partial_t^2 w(x,t)&=&2(cosh(\partial_x)-1) V^{\prime}(w(x,t))\nonumber \\&-&2 J \sum_{m=1}^{\infty}
\frac{1-cosh(m \partial_x)}{m^s} w(x,t)+\nu_{hy} \partial_x^2 \partial_t
w(x,t)\nonumber \\&+&\sqrt{D^{hy}} \partial_x^2\xi(x,t)~~.
\label{CABONOdisceq1}
\eae
The further treatment of the system depends on the value of $s$. In \cite{mingaleev98} it is
shown that in the case $s>5$ the CA yields a Boussinesq equation similar to the
result without LRI but with a different dispersion parameter. For $3<s\le 5$ it was proved that the soliton tails are no longer exponential but
algebraic and that for $s \le 3.5$ an energy gap between the soliton states and the plane wave spectra appears. For $s=4$, the equation of motion becomes a Hilbert-Boussinesq equation which can be reduced to the
integrable Benjamin-Ono form, which has algebraic soliton solutions.
\\
Since the soliton equation in \cite{brunhuber06i} was of Boussinesq-type and covered the
limit of nearest neighbour interactions, we do not
expect fundamental new effects for $s>5$. In order to investigate the behaviour of solitons with algebraic tails in the presence of thermal fluctuations, we 
restrict ourselves to the case $s=4$ where the Hilbert-Boussinesq equation and the resulting Benjamin-Ono-type solitons present a promising system to achieve
analytical results. The perturbed Hilbert-Boussinesq equation 

\eas
\partial_t^2w-c^2\partial_x^2w-\lambda \partial_x^4w-\frac{J \pi}{6}
\mathcal{H}(\partial_x^3 w)+2 \partial_x^2 w^2 =\nonumber \\
=\nu_{hy} \partial_x^2 \partial_t u+ \sqrt{D^{hy}} \partial_x \xi(x,t)
\label{HB}
\eae
with $c^2=(1+J \pi^2/6)$ und $\lambda=(1/12-J/24)$ 
follows from expanding in (\ref{CABONOdisceq1}) the operators for the
nearest-neighbour part and for the long-range part $Q(s,\partial_x)=2\sum_m  (1-cosh(m \partial_x))/m^s$
\eas
2 (cosh(\partial_x)-1)=2 \bigg(
\frac{\partial_x^2}{2!}+\frac{\partial_x^4}{4!}+... \bigg)
\\
Q(4,\partial_x)=-\frac{\pi^2}{6} \partial_x^2-\frac{\pi}{6}
\mathcal{H}(\partial_x^3)+\frac{1}{24} \partial_x^4,
\eae
where $\mathcal{H}(f(x))=1/\pi P \int_{-\infty}^{\infty} f(y)/(y-x) dy$  denotes the Hilbert transform and P the Cauchy principal value. Similarly like in
\cite{brunhuber06i}, we will have to rewrite the soliton equation (\ref{HB}) in absolute displacement coordinates in order to find its Lagrangian density (in the case $\nu_{hy}=0$).
Notice that the $\lambda$-term was neglected in reference \cite{mingaleev99}.

\begin{widetext}
\begin{figure}[htbp]
\caption{(Color online) Soliton solutions for power-law ($s=4$) and Kac-Baker long-range interactions.
The two solitons in the left and the right panel each have the same velocity $c_o$ and the same energy $H$.}
\hspace*{-0.5cm}\subfigure{\includegraphics[width=12cm,angle=270]{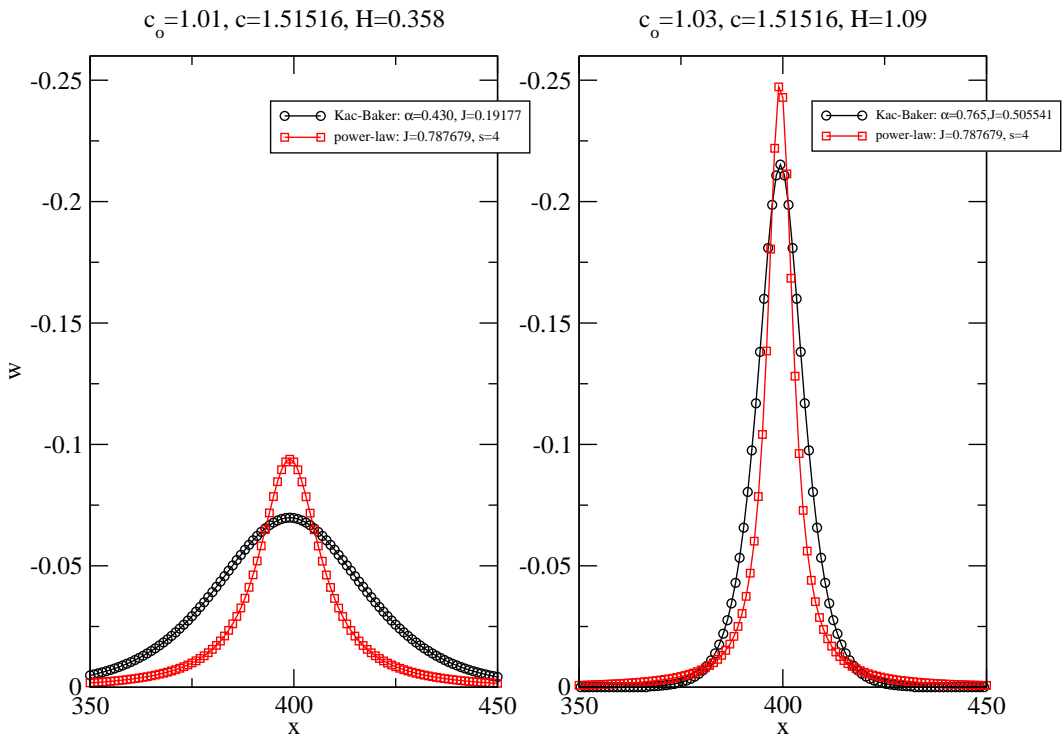}}
\label{power-law-Kac-Baker}
\end{figure}
\end{widetext}
\section{Collective Variables}
The Hilbert-Boussinesq equation in absolute displacement coordinates $u(x,t)$

\eas
\partial_t^2u-c^2\partial_x^2u-\lambda \partial_x^4u-\frac{J \pi}{6}
\mathcal{H}(\partial_x^3 u)+2 \partial_x u \partial_{xx}u=\nonumber \\
=\nu_{hy} \partial_x^2 \partial_t u+ \sqrt{D^{hy}} \partial_x \xi(x,t)
\label{BONOdgl}
\eae
can be derived for $\nu^{hy}=0$ from the Lagrangian density 
\eas
\mathcal{L}&=&\frac{u_t}{2}-\frac{c^2 u_x^2}{2}+\frac{u_x^3}{J \pi}+\frac{J
\pi}{6} u_{xx} \mathcal{H}(u_x)+\lambda \frac{u_{xx}^2}{2}\nonumber \\&-&\sqrt{D^{hy}} u_x \xi(x,t)~~.
\eae

As a CV ansatz with $X(t)$ and $\sigma(t)$, we will use the Benjamin-Ono-type soliton shape 
\eas
u(x,t)=-A_o Arctan[\sigma(t) (x-X(t))]
\nonumber \\
w(x,t)=-\frac{A_o
\sigma(t)}{1+\sigma(t)^2 (x-X(t))^2} ~.
\label{ansatz}
\eae
One can check that this ansatz yields the correct values
\eas
A_o=\frac{J \pi}{3}+4\lambda \sigma_o~,~\sigma_o=\frac{6}{\pi J} (v^2-c^2)
\eae
(compare with \cite{mingaleev99}) when one minimizes the action for  
\eas
L&=&\int dx \mathcal{L}=\nonumber \\
&=&\frac{A^2 \pi}{4} \bigg(\frac{\dot{\sigma}^2}{\sigma^3}+\sigma \dot{X}^2
\bigg) -\frac{c^2 A^2 \pi \sigma}{4}-\frac{\pi}{8} A^3 \sigma^2+\frac{A^2 J
\Pi^2 \sigma^2}{48}\nonumber \\&+&\frac{A^2 \pi \gamma \sigma^3}{8}
+\sqrt{D^{hy}} \int dx \frac{A
\sigma}{1+\sigma^2(x-X(t))^2} \xi(x,t)~~.
\label{BONOL}
\eae
(with $\lambda=0$ and $\sqrt{D^{hy}}=0$) for the soliton parameters $\sigma$, $A$ and $X(t)$ assuming a coherent excitation with constant velocity ($\dot{\sigma}=\dot{A}=0$, $\dot{X}=v$).  
\\

To include the damping, we proceed like in \cite{brunhuber06i} with the generalized Hamilton principle of Ostrovsky et al. \cite{ostrovsky71,jeffrey}
\eas
\frac{\delta <\mathcal{L}>}{\delta X_i}=\frac{\partial <\mathcal{L}>}{\partial X_i}-\frac{\partial
}{\partial t} \frac{\partial <\mathcal{L}>}{\partial \dot{X}_i}=- \bigg< \Phi
u_{i} \bigg>
\label{Leq}
\eae
where $\Phi=\nu^{hy} u_{xxt}$ is a dissipative field and the brackets signify a spatial integration over $x$.
\\
In Fig. \ref{power-law-Kac-Baker}, we show two solitons with $c_o=1.01$ (left panel) and $c_o=1.03$ (right panel) on FPU chains with power-law LRI ($s=4$,
$J=0.7877$, $c=1.51516$) and for the Kac-Baker LRI (described in \cite{brunhuber06i}). The values of $\alpha$ and $J$ for the solitons with the exponential tails
in the Kac-Baker case were chosen to produce a solution with the same velocity and soliton energy $H$. The power-law LRI yield solutions with lorentzian shape (algebraic
solitons)
whereas the Kac-Baker LRI yield solitons which are well approximated by a $sech$ (exponential solitons).

\section{Langevin system}

The eqs. (\ref{Leq}) together with the ansatz (\ref{ansatz}) leads to following
stochastic equations for the collective variables $\sigma$ and $X$:
\eas
\dot{\sigma} \dot{X}=\frac{4 \sqrt{D^{hy}} \sigma^3}{A_o \Pi} \int dx
\frac{\bar{x} \xi(x,t)}{(1+\sigma^2 \bar{x}^2)^2}-\frac{\nu_{hy}}{2} \dot{X}
\sigma^3 \nonumber \\
(\dot{X}^2-c^2)=\frac{ A_o \sigma}{2}+\frac{ \lambda \sigma^2}{2}-\frac{8
\sqrt{D^{hy}} \sigma^2}{\pi A_o} \int dx \frac{\bar{x}^2 \xi(x,t)}{(1+\sigma^2
\bar{x}^2)^2} \nonumber \\
\label{BONOXnoisy}
\eae
with
\eas
A_o=\frac{J \pi}{3}+4 \lambda \sigma_0~,~\bar{x}=x-X(t)~~.
\eae
In the previous step, we neglected small terms $\sim \dot{\sigma}^2$,
$\sim \ddot{X}$ and $\ddot{\gamma}$ in order to obtain technically less
comprehensive stochastic equations. The justification of this step was checked
by numerical simulations of the complete system and was already successfully
applied in the case of Kac-Baker LRI \cite{brunhuber06i}.

We want to rewrite the second equation as an equation for $\dot{X}$. We expand the appearing square root for the small value $\sqrt{D^{hy}}$
to first order and use that the velocity of the damped soliton 
\eas
v_d=\sqrt{c^2+\frac{A_o}{2} \sigma(t)+\frac{\lambda}{2 } \sigma(t)^2}
\label{vdexakt}
\eae
which can be approximated by $v$, the start velocity of the soliton (because $c<v_d<v$, $c\sim v$).

We end up in a system of stochastic integro-differential equations for the CV:

\eas
\left( \begin{array}{c} \dot{\sigma} \\ \dot{X} 
\end{array} \right)  = 
\left( \begin{array}{c} A_1\\ A_2
\end{array} \right) + \int_{-\infty}^{\infty} dx \left( \begin{array}{cc}  
B_{11}&0 \\ 0&B_{22}  \end{array} \right)
\left( \begin{array}{c} \xi \\ \xi \end{array} \right)
~,
\label{sideq}
\eae
with
\eas
A_1=-\frac{\nu_{hy}}{2} \sigma^3 ~~, ~~A_2=v_d
\\
B_{11}=\frac{4 \sqrt{D^{hy}} \sigma^3}{\pi A_0 v} \frac{(x-X(t))}{(1+\sigma^2 (x-X(t))^2)^2}
\\
B_{22}=-\frac{4 \sqrt{D^{hy}} \sigma^2}{\pi A_0 v} \frac{(x-X(t))^2}{(1+\sigma^2 (x-X(t))^2)^2}~~.
\eae
We proceed like in \cite{arevalo03,brunhuber04} in order to find a statistically equivalent
Langevin system to (\ref{sideq})
(with the same
Fokker-Planck equation in the Stratonovich interpretation \cite{konotop}) which is more convenient for further numerical and analytical studies. The
Langevin-system with two independent Gaussian white noise processes reads: 
\eas
\left( \begin{array}{c} \dot{\sigma} \\ \dot{X} 
\end{array} \right) =\left( \begin{array}{c} a_1\\ a_2
\end{array} \right)+\left( \begin{array}{cc}  
b_{11}&0 \\ 0&b_{22}  \end{array} \right)
\left( \begin{array}{c} \xi^1 \\ \xi^2 \end{array} \right)
\label{langevin}
\eae
with
\eas 
a_1=-\frac{\nu_{hy} \sigma^3}{2}-\frac{D^{hy} \sigma^2}{4 A_o^2 \pi v^2} ~,~a_2=v_d
\nonumber \\
b_{11}=\frac{\sqrt{D^{hy}} \sigma^{\frac{3}{2}}}{A_o \sqrt{\pi} v}
~,~
b_{22}=\frac{
\sqrt{D^{hy}}}{ \sqrt{\sigma} A_o \sqrt{\pi} v}~.
\eae  
The corresponding system in \cite{brunhuber06i} depends similarly  
on the soliton parameters inverse width, $A_o$, velocity $v$ and on the temperature $\sqrt{D}$. Only numerical constants and the velocity
$v_d$ are different. The broadening of the soliton due to the damping
follows the same manner as for exponential solitons in \cite{arevalo03,brunhuber06i}. The result (\ref{langevin}) manifests the statement made in \cite{brunhuber06i}, that
the LRI and the resulting soliton shape determine the velocity $v_d(\sigma(t))$ of the damped soliton.

\section{small-noise expansion}

We proceed with a small-noise expansion for the parameter $\sqrt{D^{hy}}$
like in \cite{brunhuber06i,arevalo03}  in order to derive an analytical expression for the
soliton diffusion. The Langevin system (\ref{langevin}) is formally very similar
to the sytem in \cite{brunhuber06i} for Kac-Baker LRI, 
only numerical constants and the soliton parameters are different. 
For algebraic solitons, we can approximate (\ref{vdexakt}) with
$v_d=c+A_o/4c~ \sigma(t)$ because of the smallness of $\lambda$. The analytical
solution in this case is technically less extensive and leads practically to the
same results. The approximation for $v_d$ is not possible in the case of Kac-Baker LRI
in \cite{brunhuber06i} because the long-range forces yield a larger term which
is quadratic in the inverse soliton width.  
\\
The result in the first order of $\sqrt{D^{hy}}$ reads (see Appendix)
\begin{widetext}
\eas
Var[X(t)]=D^{hy} \bigg(\frac{A_o^2 [ -16 (-1+\sqrt{\eta})+ (\eta-1) (18-10
\sqrt{\eta}+(\eta-1)(\eta+2))]}{60 A_o^2 c^2 \nu_{hy}^3\pi \sigma_o^3 v^2 \eta^{\frac{3}{2}
}}-
\frac{2c^2 \nu_{hy}^2 \eta(-1+\sqrt{\eta}-(\eta-1) (\eta+1))}{3 A_o^2 c^2
\nu_{hy}^3\pi \sigma_o^3 v^2 \eta^{\frac{3}{2} }} \bigg) \nonumber \\
\label{snexp}
\eae
\end{widetext}
with
\eas
\eta=1+t_{r}=1+\sigma_o^2 \nu_{hy} t~~.
\label{tr}
\eae
The time-dependence of the position variance depends on the time scale $t_r$,
which explains
that the large time limit (and the superdiffusive behaviour) sets in earlier for
high-velocity solitons (large values of $\sigma_o$) for the same damping constant. 
The corresponding result for Kac-Baker LRI is even more complex, which stems from
the longer expression for $v_d$ \cite{brunhuber06i}. The time scale which describes the
broadening of the exponential soliton and the diffusion is $t_r=1.6 \nu_{hy}
\gamma_o^2$ where $\gamma(t)$ is the inverse width of the exponential soliton  
\\
We are interested in the soliton diffusion for large times $t_{r}$. Substituting
(\ref{tr}) into (\ref{snexp}) and looking for the leading order of $t_r$ yields:
\eas
Var[X(t)]_{inf}\approx \frac{D^{hy}}{60 c^2 \nu_{hy}^{1.5} \pi  v^2} t^{1.5}~,
\label{varianz}
\eae
which differs from the result in \cite{brunhuber06i} only by the numerical
constant.
\\
For very small times, the result (\ref{snexp}) gives a linear time-dependence 
\eas
Var[X(t)]_0 =\frac{4}{3}\frac{D^{hy}}{A_o^2 \pi \sigma_o^2 v^2} t ~.
\label{normalV}
\eae 
The results of the small-noise expansion shows the same characteristic dependencies on the soliton parameters as for Kac-Baker LRI in \cite{brunhuber06i}. 
For intermediate times, the result (\ref{snexp}) describes the transition when
the position variance turns from the small-time dependence into the long-time
limit wherefore a stronger increase in time (mainly $\sim t^2$) can be
observed.


\section{Simulations}
The simulation of system (\ref{BONOdisceq1}) was performed in the same manner as described
in \cite{brunhuber06i} but with power-law long-range coupling instead of the 
Kac-Baker LRI. The position variance of the soliton was calculated from 100
different realizations of the chain. We used the solitons depicted in Fig. \ref{power-law-Kac-Baker}
 as initial conditions of the chain. The soliton with $c_o=1.01$ ($\sigma_o\approx0.1$) is much broader than the lattice spacing whereas the
soliton with $c_o=1.03$ ($\sigma_o \approx 0.3$) is rather discrete in the soliton center. The coupling parameter is always fixed at the value $J=0.7877$ which yields the
sound velocity $c=1.51516$, the same value was used in \cite{brunhuber06i}. In order to compare the results of the Kac-Baker LRI with the power-law LRI and to be able to draw
meaningful conclusions , we also use the same damping constant $\nu_{hy}=0.01$ and the same temperature $T=0.0001$ as in Ref. \cite{brunhuber06i}.    

\begin{figure}[htbp]
\caption{(Color online) Position variance of a low-velocity ($1.01$) and a high-velocity ($1.03$) soliton on a anharmonic chain with power-law long-range interactions
($J=0.7877$, $s=4$). We compare the simulation results with equation (\ref{snexp}), the result of a small-noise expansion of the Langevin system (\ref{langevin}) of the CV
theory.}
\hspace*{-0.5cm}\subfigure{\includegraphics[width=7cm,angle=270]{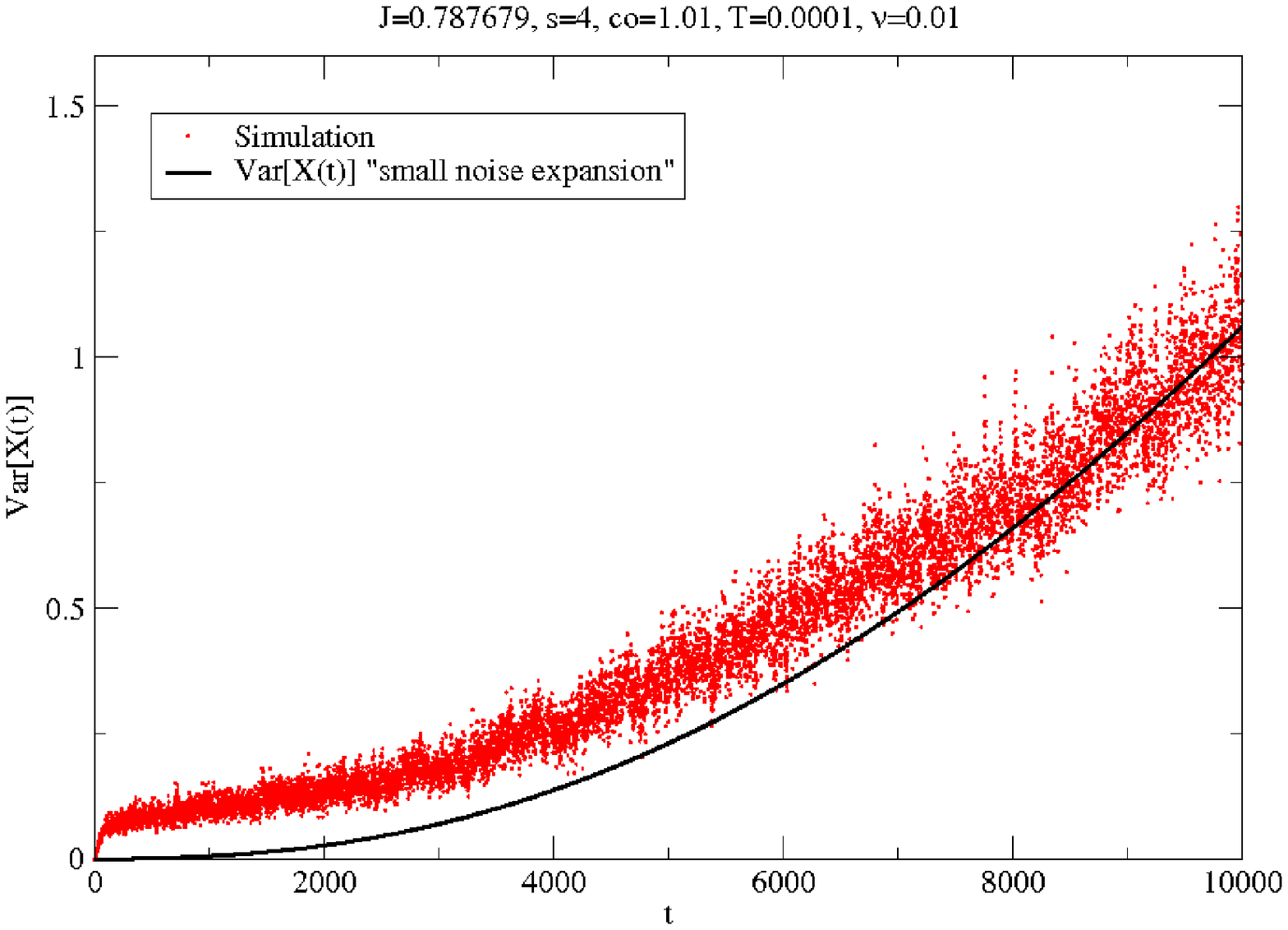}}
\hspace*{-0.5cm}\subfigure{\includegraphics[width=7cm,angle=270]{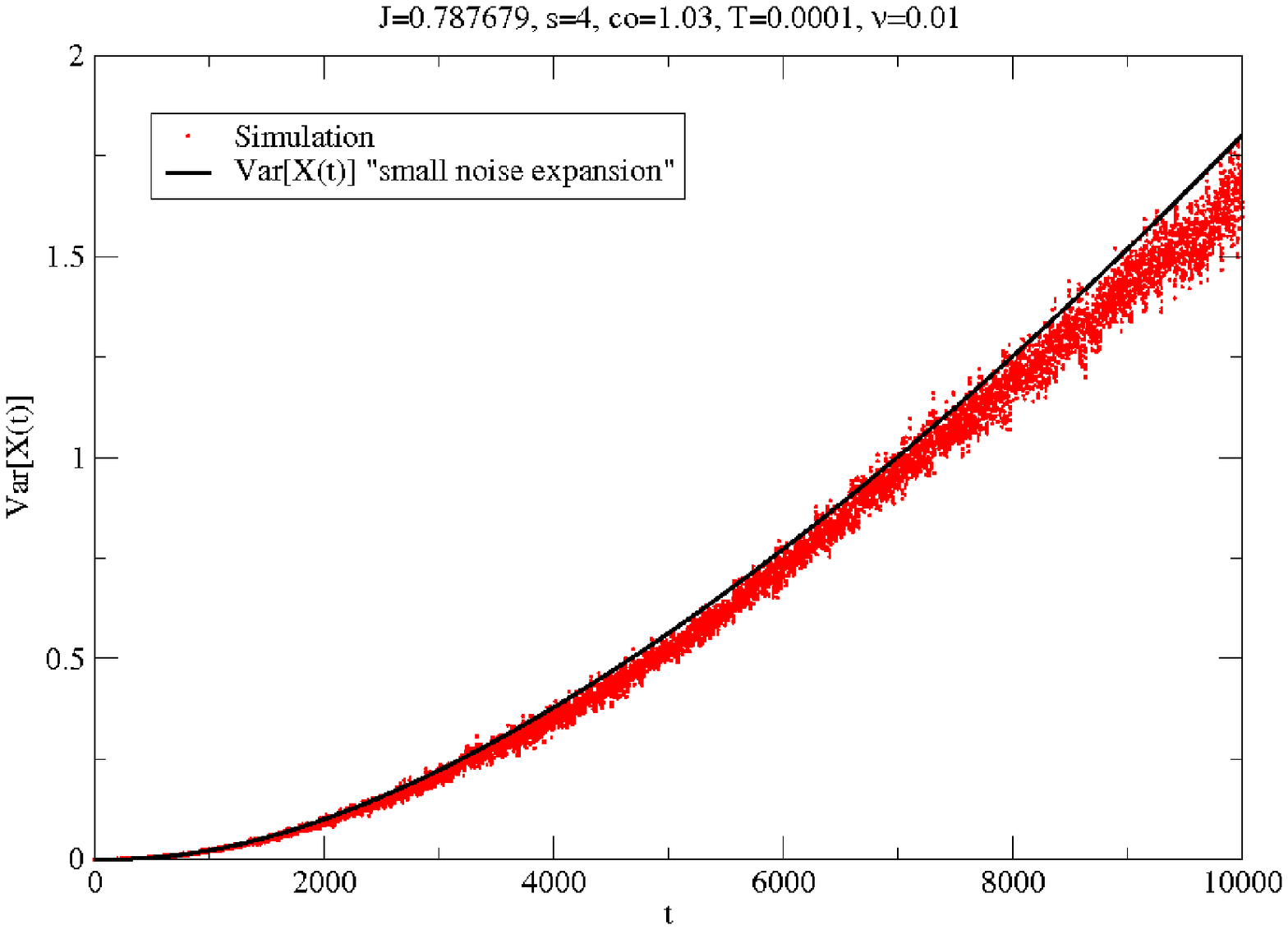}}
\label{diff}
\end{figure}
The analytical result for the position variance (\ref{snexp}) agrees rather well
with the simulation results (Fig. \ref{diff}). 
In general, one expects the analytical results to be slightly smaller than the simulation
results because the possible contribution of phonons is not regarded in the calculations of Section III. 
It was shown for a related short-range model that the phonons have
indeed a perceptible influence on the soliton diffusion \cite{mertens05}.
The analytical result for the low-velocity soliton diffusion for $c_o=1.01$ is
too small for small times. Similar discrepancies appeared for Kac-Baker LRI in 
\cite{brunhuber06i} and they were interpreted as the influence of phonons which cause
higher values in the simulations. 
The analytical result for the high-velocity soliton $c_o=1.03$ deviates from the
simulation results for large times. This feature seems to stem from our restriction to 
first order perturbations in the small-noise expansion. When the solitons are
very narrow, the width of the soliton changes quite fast and the first-order
correction of $\sigma(t)$ can fail. Such deviations were also observed for FPU
chains with or without Kac-Baker LRI \cite{brunhuber06i,arevalo}.    
\\
The analytical result for the soliton diffusion depends only on the time scale $t_r=\nu_{hy} \sigma_o^2 t$ and for large times the diffusion should approach the result
(\ref{varianz}) from below. Since the time scale $t_r$ depends on the width of the soliton, the long-time approximation for the position variance for high-velocity
solitons will be valid for smaller values of $t$ than for low-velocity solitons. For $c_o=1.03$ ($\sigma_o\approx0.3$) $t_r>1$ is valid for $t>1111$ whereas the corresponding time
for $c_o=1.01$ ($\sigma_o \approx0.1$) is $t>10000$. Therefore, we will check the validity of the long-time approximation result of the position variance for the soliton
with $c_o=1.03$. In deriving the long-time limit (\ref{varianz}) of the position
variance, we recognized that the approximation 
\eas
Var[X(t)]_{h} \approx \frac{D}{60 c^2 \nu_{hy}^2 \pi \sigma_o^3 v^2}
\frac{t_r^3}{t_r^{1.5}+1.5 \sqrt{t_r}}
\label{interim}
\eae
describes the simulation results very well for all times $t$. Neglecting the $\sqrt{t_r}$-term for large $t_r$ yields immediately the result (\ref{varianz}).

\begin{figure}[htbp]
\caption{(Color online) Position variance of a high-velocity soliton ($c_o=1.03$) on an anharmonic chain with power-law long-range interactions
($J=0.7877$, $s=4$). The simulation and two different long-time approximations for (\ref{snexp}), $Var[X(t)]_{inf}$ (\ref{varianz}) and $Var[X(t)]_{h}$ (\ref{interim}), are
shown. For $t=10000$ ($t_{r}\approx9$), the long-time limit is not yet reached, but the simulation results approach this function for long times. }
\subfigure{\includegraphics[width=7cm,angle=270]{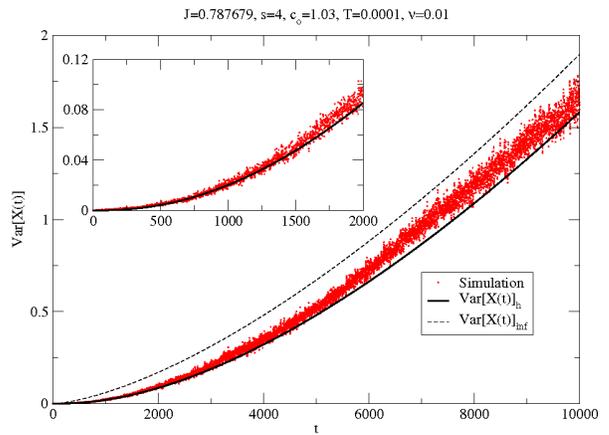}}
\label{BONOdiff}
\end{figure}

The comparison between the simulation results and $Var[X(t)]_h$ or $Var[X(t)]_{inf}$ in Fig. \ref{BONOdiff} shows that there is practically no linear dependence on $t$ similarly as for high-energy
solitons with exponential tails with Kac-Baker LRI in \cite{brunhuber06i}. The superdiffusion with stronger-than-linear terms in the time-dependence of the position variance is
the dominating mechanism for high-energy solitons.
The time-dependence of $Var[X(t)]_{h}$ describes the simulation results very well for all times and the soliton can be described by a time-dependence according to (\ref{varianz}) ($\sim
t^{3/2}$) for large
times.
For lower soliton energies like for $c_o=1.01$ (Fig. \ref{diff}) the normal contributions on $Var[X(t)]$ are stronger and a comparison with  $Var[X(t)]_{inf}$ yields a  
worse agreement because in this case the limit of large
$t_r$ can not be used for $t<10000$. 
\\
In the limit of short times ($t_r <<1$), the quality of the result for the linear contribution in time (\ref{normalV}) is very similar to the Kac-Baker case and 
yields values for $Var[X(t)]$ which are smaller than the simulation results. For the soliton with $c_o=1.01$ the slope of (\ref{normalV}) is approximately 
half of the value one would get from a linear fit of the simulation results for small times.

\section{Comparison} 
\begin{figure}[htbp]
\caption{(Color online) Simulation results for the position variance of two solitons, one on an anharmonic chain with power-law long-range interactions
($J=0.7877$, $s=4$) and the other soliton on an anharmonic chain with Kac-Baker
long-range interactions for $T=0.001$. The interaction radius and the coupling for the Kac-Baker case
were chosen in such a way that the two solitons have the same velocity $c_o=1.03$ and the same energy. In order to demonstrate the difference in the time-dependence for the
two solitons, we added a best fit for the two simulation results for $Var[X(t)]$. One can clearly see that the position variance of the algebraic soliton approaches a $\sim t^{1.5}$  
time-dependence, whereas in the Kac-Baker case, the time-dependence can be fitted by a quadratic and a cubic term.}
\subfigure{\includegraphics[width=7cm,angle=270]{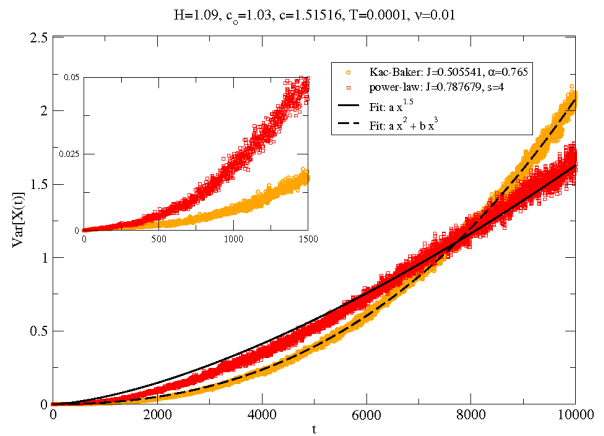}}
\label{comparison1}
\end{figure}

Up to now, we have mentioned the similarities between the thermal diffusion of solitons with algebraic and exponential tails on FPU-like chains. Now, we want to demonstrate the
main difference, namely the different time-dependence of the superdiffusion in the two cases. We want to directly compare the results for $Var[X(t)]$ and their time-dependence
for two solitons with the same energy and velocity ($c_o=1.03$) but one on a chain with power-law LRI and the other on a chain with Kac-Baker LRI. For the two solitons, normal diffusion
contributions are negligible and the best fits in Fig. \ref{comparison1} demonstrate that the position variance of algebraic solitons increases with $\sim t^{3/2}$ whereas for the $sech$-shaped solitons,
the increase goes with a quadratic ($\sim t^2$) and a cubic ($\sim t^3$) term.   
The small-noise expansion results for
both soliton types have the same long-time behaviour $(\sim t^{3/2})$ (compare
with \cite{brunhuber06i}) which is
reached by the algebraic soliton in Fig. \ref{comparison1} for $t>8000$, whereas the exponential soliton
is for times $t<10000$ still in the regime where the long-time limit is
not applicable.

\begin{figure}[htbp]
\caption{(Color online) $Var[X(t)]$ and $Var[X(t)]_{inf}$ from a small-noise
expansion for the two solitons of Fig. \ref{comparison1}. 
The algebraic soliton reaches the long-time limit $Var[X(t)]_{inf}$ much sooner
than the exponential soliton with the same initial energy $H=1.09$ and velocity
$c_o=1.03$. The result $Var[X(t)]_{inf}$ shows the same time-dependence but the
diffusion constant for algebraic solitons is approximately a factor of $\pi$
smaller. 
}
\subfigure{\includegraphics[width=7cm,angle=270]{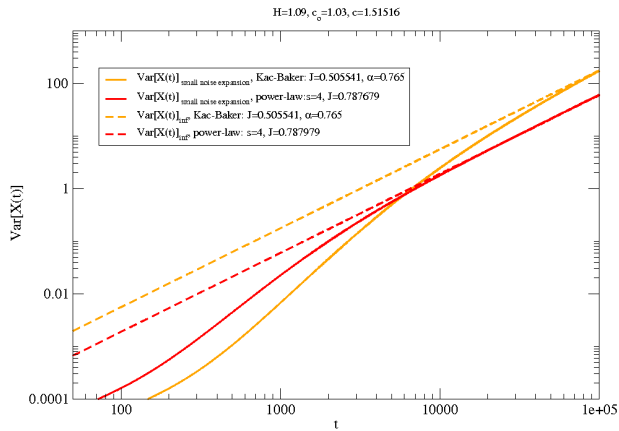}}
\label{longtime}
\end{figure}

The Langevin systems for both soliton types are identical apart from the
numerical constants and the values of the soliton parameters like the width. For Kac-Baker LRI,
the inverse soliton width $\gamma_o=0.132$ at $t=0$ is distinctly smaller than
for the power-law case with $\sigma_o=0.279$ for the same energy and soliton
velocity. Since the time scale $t_r$ depends
quadratically on the inverse width, it is not surprising that the long-time
limit for exponential solitons is valid after much longer times. This argument
is decisive for different solitons of the same type. It follows directly from the small-noise
expansion that the long-time limit for the algebraic soliton with
$c_o=1.01$ appears for $\sigma_o^2({c_o=1.03})/\sigma_o^2({c_o=1.01}) \approx
5.44$ longer times than for the algebraic soliton with $c_o=1.03$.   
\\
In order to estimate when the exponential soliton reaches the long-time
asymptotics,
we have to concentrate our studies on the discussion of the result $Var[X(t)]$ from
the small-noise expansion for algebraic and exponential solitons
(Fig. \ref{longtime}). Simulations for times much longer than $10^4$ 
would be too time-consuming. 
The double-logarithmic scales in Fig. \ref{longtime} make clear
that the Kac-Baker soliton reaches the
long-time limit $Var[X(t)]_{inf}$ after much longer times than the power-law
soliton. The time window where a stronger time-dependence than $\sim t^{3/2}$
appears is approximately ten times larger for the Kac-Baker soliton.

\begin{figure}[htbp]
\caption{(Color online) The simulation results for the mean soliton velocity 
$\langle \dot{X} \rangle$ in case of a noisy chain
with $T=0.0001$  for energetically equivalent algebraic and exponential solitons with $c_o=1.03$.
The zeroth-order results (without noise) for the soliton velocities $v_d(\sigma^{(0)}(t))$ and $v_d(\gamma^{(0)}(t))$ agree rather well
with the simulation results and demonstrate the different dynamics of the two soliton types.}
\subfigure{\includegraphics[width=7cm,angle=270]{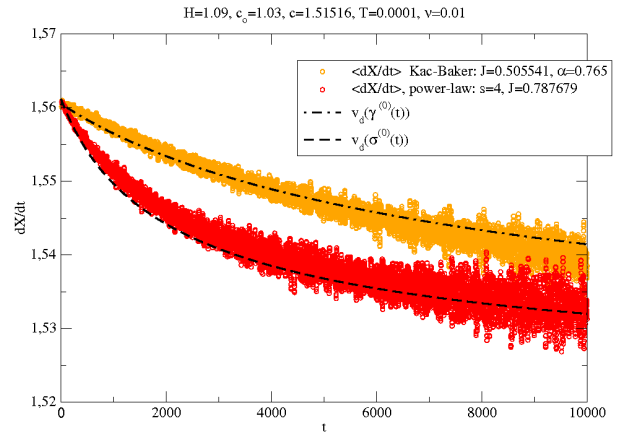}}
\label{vd}
\end{figure}

The influence of the soliton shape determines the diffusion constant in
$Var[X(t)]_{inf}$, which is approximately a factor of $\pi$ larger for
exponential solitons. One can equivalently state that the longer transient times to the long-time
limit for the exponential
soliton is 
a consequence of the higher value of the diffusion constant. This remarkable difference for the two
soliton types results only from the different soliton shapes because the
numerical constant in (\ref{varianz}) is determined by the numerical constants
in the Langevin system (\ref{langevin}) which follow from spatial integrations over the soliton
profile in the CV procedure.
\\
The time evolution of the velocity of the energetically equivalent algebraic and exponential soliton, 
$v_d(\sigma(t))$ and $v_d(\gamma(t))$, is rather different despite of the same start value
and the structural similarities for $v_d$. 
In Fig. \ref{vd}, we present the simulation results for the mean velocity
$\langle \dot{X}(t) \rangle$ of the two solitons in comparison with the zeroth order
expressions for $v_d$ from the small-noise expansion: $v_d(\sigma^{(0)}(t))$ and
$v_d(\gamma^{(0)}(t))$ \cite{brunhuber06i}.
 \\
These hold for the zero temperature case but describe the mean velocity of the two
noisy solitons already quite well. 
We can directly see that the power-law soliton approaches the long-time limit with small soliton velocities earlier than the Kac-Baker soliton. 
The slower dynamics of the Kac-Baker soliton is caused by its broad, exponential
shape and the resulting slower time scale $t_r$ which controls the  time-evolution 
of $\gamma^{(0)}(t)$. The 
dependence of $v_d$ on a term $\sim \gamma^2$ for exponential solitons 
(which is negligible for algebraic solitons) influences the soliton velocity (and therefore the 
superdiffusion), especially for pronounced LRI character and for small times when the
long-time limit is not yet reached.

\section{Conclusions}

We investigated the diffusion of algebraic solitons on FPU-like chains with power-law long-range interactions and derived analytically a result for the position variance which
agrees well with the simulations of the discrete system. The results clearly
reveal that the soliton diffusion and the damping-induced reshaping of the
soliton depends on the
time scale $t_r=\sigma_o^2 \nu_{hy}t$ which is given by the soliton width and
the damping constant. For very broad solitons, the damping effects are
quite weak because the particles in the soliton profile have
moderate relative velocities and $t_r$ evolves slowly in $t$. A small-noise expansion of the Langevin system yields an analytical result for the soliton diffusion which
depends on $t_r$ and predicts a superdiffusive long-time behaviour $\sim
t^{3/2}$. This result makes clear why the superdiffusion, which sets in for
large $t_r$, dominates the diffusion of high-energy solitons ($\sigma_o^{-1}$
small). These basic features were also found in earlier studies for solitons with
exponential shape. The superdiffusion of lattice solitons can therefore be
regarded as generic when the lifetime of the solitons is much larger than $1/\sigma_o^2 \nu_{hy}$. On
chains with NNI, this situation can
often not be reached for low-energy solitons because they are rapidly destroyed
by the fluctuations and the damping. {\it The LRI stabilize the soliton and lead to higher soliton energies and
lifetimes which makes the superdiffusion the dominating mechanism}. 
\\
The shape of the soliton does not change the basics of the diffusion
process. The Langevin systems for the collective variabels (inverse width and
soliton position) 
are technically similar, only numerical constants and the dependence of the soliton velocity on the inverse width are different. We proved in simulations and by analyzing the
analytical result that despite of the similarities, the superdiffusive behaviour
of the two soliton types exhibits striking differences even when the soliton energy and velocity coincide. The algebraic solitons 
are generally narrower than exponential solitons which leads to faster dynamics
for the time scale $t_r$ and an earlier validity of the long-time limit with the characteristic $\sim t^{3/2}$
time-dependence. The broadness of the exponential solitons, the corresponding
slower dynamics for $t_r$ and the stronger dependence on higher orders of the inverse width in the
soliton velocity shift the $\sim t^{3/2}$ region to larger times $t$ which
leads to the observed characteristic quadratic and cubic time-dependence in the simulations.
{\it The perhaps most remarkable effect of the different soliton shapes is that
the exponential solitons show larger position variances than their algebraic
counterparts for sufficiently long times}. The diffusion constant in the long-time
limit is for algebraic solitons approximately a factor of $\pi$ smaller than for exponential solitons. This effect is only caused
by the different soliton shapes which determine numerical constants in terms which describe the coupling between the soliton velocity and the fluctuations of the soliton width.  
Further studies for different values of $s$ would be helpful to explain the influence of power-law interactions on the soliton diffusion. Especially the diffusion of
solitons on chains with power-law LRI for the case $s<3.5$
could be interesting because a gap between the plane-wave spectra and the soliton energies appears.

\appendix*
\section{Small-noise expansion}
The small-noise expansion of the Langevin set (\ref{langevin}) follows reference \cite{gardinier} and was already successfully applied to the problem of the diffusion of
low-velocity solitons in the FPU system without long-range interactions \cite{arevalo03}. We seek an asymptotic solution of the form
\eas
\sigma(t)=\sigma^{(0)}(t)+\epsilon \sigma^{(1)}(t)+ ...\\
X(t)=X^{(0)}(t)+ \epsilon X^{(1)}(t)+... 
\eae
where $\epsilon$ is a small parameter which is formally introduced to consider the influence of the noise terms as small perturbations ($\sqrt{D^{hy}}\sim \epsilon$).
The contribution of the drift term in the different orders of $\epsilon$ are calculated following the rule
\eas
a_i(\sigma(t))&=&a_i(\sigma^{(0)}+\sum_{m=1}^{\infty} \epsilon^m
\sigma^{(m)}(t))\nonumber \\&=&a_i^{(0)}(t)+\epsilon \sigma^{(1)}(t)\frac{d a_i(\sigma^{(0)}(t))}{d
\sigma^{(0)}(t)}+
... ~~.
\eae
In order to minimize the technical efforts, it is advisable to approximate the velocity $a_2=v_d$ by the following expression
\eas
a_2&=&\sqrt{c^2+\frac{A_o}{2} \sigma(t)+\frac{\gamma}{2 } \sigma^2(t)}\approx c
\sqrt{1+\frac{A_o}{2 c^2} \sigma(t)}\nonumber \\&\approx& c+\frac{A_o}{4 c} \sigma(t)~~,
\eae
which was checked to be appropriate for typical soliton parameters $J$, $v$.
The equations in the order $\epsilon^0$ read:
\eas
d \sigma^{(0)}(t)=-\frac{\nu_{hy}}{2} \sigma^{(0)}(t)^3 dt
\\
d X^{(0)}(t)=\bigg( c+\frac{A_0}{4 c} \sigma(t) \bigg) dt~~.
\eae
These equations describe the damped soliton. The damping induced broadening of the soliton is formally equivalent to the case of exponential soliton solutions in
\cite{brunhuber06i}. In the expression for the soliton velocity it is obvious that the
damped soliton gets broader (smaller $\sigma$) which leads to a slowdown of the
soliton. In the limit of long times, the soliton width diverges and the soliton approaches the velocity of sound.
\\
The first order corrections due to the noise read:
\eas
d \sigma^{(1)}(t)=-\sigma^{(1)}(t)\frac{3}{2} \nu_{hy} \sigma^{(0)}(t)^2)
dt+\frac{\sqrt{D} \sigma^{(0)}(t)^{\frac{3}{2}}}{A_0 \sqrt{\pi} v} dW_1\nonumber \\
\\
dX^{(1)}(t)=\sigma^{(1)}(t) \frac{A_0}{4 c}   
dt+\frac{ \sqrt{D}}{ \sqrt{\sigma^{(0)}}\sqrt{\pi} A_0 v} dW_2
\label{x1}
~~.\eae
If we substitute the result for $\sigma^{(0)}=\sigma_o/\sqrt{\nu^{hy} \sigma_o^2 t+1}$ into (\ref{x1}) we calculate the 
first-order expression for the dislocation of the soliton due to the noise
\eas
X^{(1)}(t)&=&\frac{\sqrt{D} \sigma_o^(\frac{3}{2})}{4 c\sqrt{\pi} v} \int_{0}^{t} dt{\eta^{\prime}}^{-\frac{3}{2}}  \int_{0}^{t^{\prime}} dt^{\prime \prime}
{\eta^{\prime \prime}}^{\frac{3}{4}} d
W_1(t^{\prime \prime}) \nonumber \\&+& \frac{\sqrt{D}}{ \sigma_o \sqrt{\pi} A_o v} \int_{0}^{t}
{\eta^{\prime}}^{\frac{1}{4}} dW_2(t^{\prime})~~.
\\
\eta^{\prime}&=&1+\nu_{hy}\sigma_o^2 t^{\prime}
\eae
After some straightforward calculations we can calculate the position varinace 
\eas
Var[X_1(t)]=\lim_{s \rightarrow t} \rangle X_1(t) X_1(s) \langle
\eae

which yields the result (\ref{snexp}) where we can finally set $\epsilon=1$ which is justified if the noise in the system is sufficiently small.

\end{document}